\documentclass[conference]{IEEEtran}
\usepackage{amsmath, amssymb, bm, cite, epsfig, psfrag}
\usepackage{graphicx}
\usepackage{soul} 
\usepackage[top    = 0.7in,
bottom = 0.8in,
left   = 0.625in,
right  = 0.625in]{geometry}

\usepackage{dblfloatfix}
\usepackage{array}
\usepackage{lipsum}

\newcolumntype{P}[1]{>{\centering\hspace{0pt}}p{#1}}
\newcolumntype{M}[1]{>{\centering\hspace{0pt}}m{#1}}
\newcolumntype{L}{>{\centering\arraybackslash}m{3cm}}
\usepackage[font=small]{caption}
\usepackage{epstopdf}	
\usepackage{longtable}
\usepackage{supertabular,booktabs}
\usepackage{bbm}
\usepackage{multirow}
\usepackage[usenames,dvipsnames]{xcolor}

\usepackage{subfigure}
\usepackage{etoolbox}
\usepackage{pbox}
\usepackage{fixltx2e}
\usepackage{tabu}	
\usepackage{filecontents}
\usepackage{enumerate}
\usepackage{textcomp}

\usepackage{colortbl}
\usepackage{fancyhdr}

\usepackage{bm}

\newtoggle{conference}

\togglefalse{conference} 
\graphicspath{{figures/}}

\newcolumntype{?}{!{\vrule width 2pt}}

\fancypagestyle{firststyle}{
	\fancyhf{}
	\fancyhead[L]{Y. Xing, F. Hsieh, A. Ghosh, and T. S. Rappaport, ``High Altitude Platform Stations (HAPS): Architecture and System Performance,''  2021 \textit{IEEE 93rd Vehicular Technology Conference (VTC-Spring)}, April 2021, pp. 1-6.}     

}
\IEEEoverridecommandlockouts

\begin{document}
\title{High Altitude Platform Stations (HAPS): \\ Architecture and System Performance}
\author{\IEEEauthorblockN{Yunchou Xing\IEEEauthorrefmark{1}, Frank Hsieh\IEEEauthorrefmark{2}, Amitava Ghosh\IEEEauthorrefmark{2}, and Theodore S. Rappaport\IEEEauthorrefmark{1}}
	
	\IEEEauthorblockA{	\small \IEEEauthorrefmark{1}NYU WIRELESS, NYU Tandon School of Engineering, Brooklyn, NY, 11201, \{ychou, tsr\}@nyu.edu\\
		\IEEEauthorrefmark{2}NOKIA Bell Labs, Naperville, IL, 60563, \{frank.hsieh, amitava.ghosh\}@nokia-bell-labs.com }
	\vspace{-0.7cm}
	
	\thanks{This research is supported by the NYU WIRELESS Industrial Affiliates Program, Nokia Bell Labs, and National Science Foundation (NSF) Research Grants: 1909206 and 2037845.}
}

\maketitle
\thispagestyle{firststyle}

\begin{abstract} 
High Altitude Platform Station (HAPS) has the potential to provide global wireless connectivity and data services such as high-speed wireless backhaul, industrial Internet of things (IoT), and public safety for large areas not served by terrestrial networks. A unified HAPS design is desired to support various use cases and a wide range of requirements. In this paper, we present two architecture designs of the HAPS system: i) repeater based HAPS, and ii) base station based HAPS, which are both viable technical solutions. The energy efficiency is analyzed and compared between the two architectures using consumption factor theory. The system performance of these two architectures is evaluated through Monte Carlo simulations and is characterized in metrics of spectral efficiency using LTE band 1 for both single-cell and multi-cell cases. Both designs can provide good downlink spectral efficiency and coverage, while the uplink coverage is significantly limited by UE transmit power and antenna gain. Using directional antennas at the UEs can improve the system performance for both downlink and uplink. 
     
\end{abstract}


\section{Introduction}~\label{sec:intro}

High altitude platform station (HAPS) is a communication platform deployed in the stratosphere (e.g., 18-24 km above the ground), which can utilize solar power to operate for several months without disturbances and to provide connectivity for a large area (e.g., with a diameter of 200 km) \cite{kara05HAPS}. HAPS experiences a low propagation delay (e.g., 1-2 ms) and negligible Doppler shift compared to low earth orbit (LEO) satellites, and it is much easier and cheaper to launch and maintain \cite{frank19b}. HAPS can support various applications including mobile broadband in rural areas, Internet of Things (IoT), public safety, autonomous factory, and disaster relief \cite{mohammed11HAPS,frank19a,rappaport19access,xing21a}. There are many ongoing HAPS projects and innovations such as Google Loon, HAPSMobile, and Airbus Zephyr to provide 4G LTE and 5G services to the sparsely populated areas in the global wireless connectivity.

The \textcolor{black}{global cellular industry standardization organization}, 3GPP, is working on specifying 5G new radio (NR) standard for non-terrestrial networks (NTN), including \textcolor{black}{geostationary earth orbit} (GEO) satellites, LEO satellites, HAPSs, and unmanned aerial vehicles (UAVs), and the initial specification is expected to be delivered in 3GPP NR Release 17 \cite{3GPP38.811}. Extensive literature surveys of air-to-ground communication channel modeling in the frequency range of 1-5 GHz with an altitude up to 11 km are provided in \cite{khawaja19survey,Khuwaja18survey,amorim17UAV} including various scenarios of urban, suburban, rural, hilly, forest, and ocean. \textcolor{black}{Air-to-ground and air-to-air communications utilizing millimeter wave at 60 GHz with propagation measurements, channel models, and beam alignment algorithms are presented in \cite{polese20UAV,garcia20UAV}.}

In this paper, we propose two types of architectures of HAPS, where the HAPS is working as a base station (i.e., regenerative architecture) and as a repeater (i.e., bent-pipe architecture), respectively, and present their corresponding system performance. Details of the architecture designs are shown in Section \ref{sec:architecture}. System model and simulation settings are presented in Section \ref{sec:sysmodel}. System performance results of the two types of architectures for single-cell and multi-cell cases are evaluated in Section \ref{sec:results}. Finally conclusions are drawn in Section \ref{sec:conclusion}.

\begin{table*}[ht]\caption{Specific solutions of two architectures of HAPS.}\label{tab:solutions}
	\centering
	\begin{tabular}{|c|c|c|c|c|c|}
		\hline
		\textbf{Solution}   & \textbf{\begin{tabular}[c]{@{}l@{}}Payload \\ requirements\end{tabular}}                          & \textbf{\begin{tabular}[c]{@{}l@{}}Beam\\ control\end{tabular}} & \textbf{\begin{tabular}[c]{@{}l@{}}L1/MAC\\ Latency\end{tabular}}           & \textbf{\begin{tabular}[c]{@{}l@{}}FL/AL single and\\ bandwidth\end{tabular}}                              & \textbf{Notes}                                                                    \\ \hline
		Simple repeater     & \multirow{6}{*}{\begin{tabular}[c]{@{}l@{}}Light\\  \\   \\ \\ \\
				Heavy\end{tabular}} & NO                                                              & \multirow{4}{*}{\begin{tabular}[c]{@{}l@{}}Longer\\ (FL + AL)\end{tabular}} & Common                                                                                                     &                                                                                   \\ \cline{1-1} \cline{3-3} \cline{5-6} 
		Advanced repeater   &                                                                                                   & \multirow{5}{*}{Yes}                                            &                                                                             & \multirow{2}{*}{\begin{tabular}[c]{@{}l@{}}Depending on \\ solutions\end{tabular}}                         & \begin{tabular}[c]{@{}l@{}}Beam control solution is needed\end{tabular}         \\ \cline{1-1} \cline{6-6} 
		Relay station (IAB) &                                                                                                   &                                                                 &                                                                             &                                                                                                            &                                                                                   \\ \cline{1-1} \cline{5-6} 
		BS (RU only)        &                                                                                                   &                                                                 &                                                                             & \multirow{3}{*}{\begin{tabular}[c]{@{}l@{}}Independent\\ (Possible separate\\ optimizations)\end{tabular}} & \begin{tabular}[c]{@{}l@{}}Heavy data in FL due to low layer split\end{tabular} \\ \cline{1-1} \cline{4-4} \cline{6-6} 
		BS (DU + RU)        &                                                                                                   &                                                                 & \multirow{2}{*}{\begin{tabular}[c]{@{}l@{}}Shoter\\ (AL only)\end{tabular}} &                                                                                                            &                                                                                   \\ \cline{1-1} \cline{6-6} 
		BS (full BBU)       &                                                                                                   &                                                                 &                                                                             &                                                                                                            &                                                                                   \\ \hline
	\end{tabular}
	\vspace{-0.5cm}
\end{table*}   

\section{Two Types of Architectures of HAPS}~\label{sec:architecture}
HAPS stations are envisioned to support various use cases utilizing different carrier frequencies and bandwidth (e.g., LTE, cmWave NR, and mmWave NR), such as wide-coverage backhaul for non-terrestrial group mobility at $\sim10$ Gbps data rate, high-speed wireless backhaul for temporary industrial networks with around 1 Gbps data rate, and mobile broadband services and IoT for wide-area coverage (e.g., rural area connectivity or disaster relief) at tens of Mbps data rate \cite{frank19b}.

\begin{figure}[htbp]
	\centering
	\includegraphics[width=0.9\linewidth]{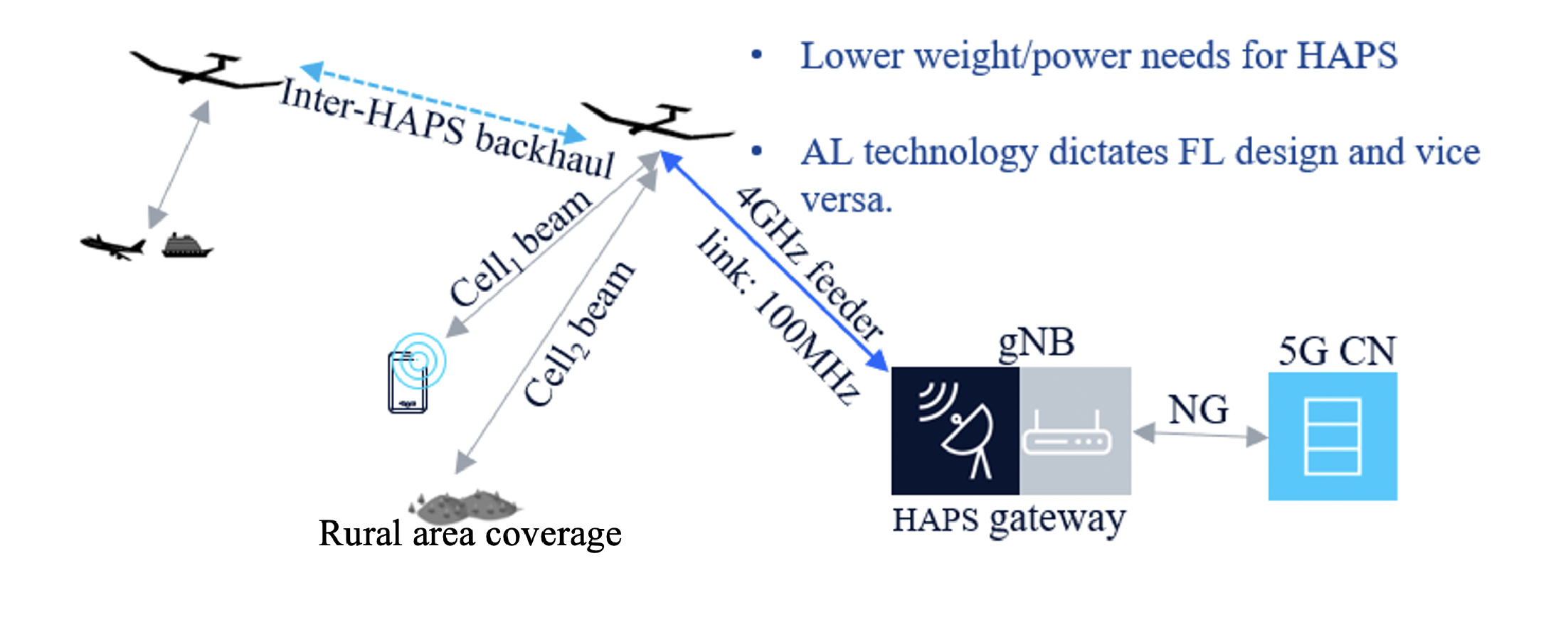}
	\caption{Bent-pipe HAPS architecture: HAPS as a repeater.}			
	\label{fig:BParch}
			\vspace{-0.1in}
\end{figure}

There are two types of architectures of HAPS that can be considered as a desired unified solution to satisfy the wide range of requirements for the various HAPS use cases. The first type is transparent or bent-pipe (BP) architecture where the HAPS station is working as an RF repeater or a relay station, as shown in Fig. \ref{fig:BParch}. The repeaters can be a simple RF repeater, an advanced repeater with beam control, or a relay station (integrated access and backhaul, IAB), each with different payload and power consumption requirements. The BP architecture requires low weight and power consumption on HAPS. However, the access link (AL) air interface and the feeder link (FL) design need to be considered together.

\begin{figure}
			\centering
			\includegraphics[width=0.9\linewidth]{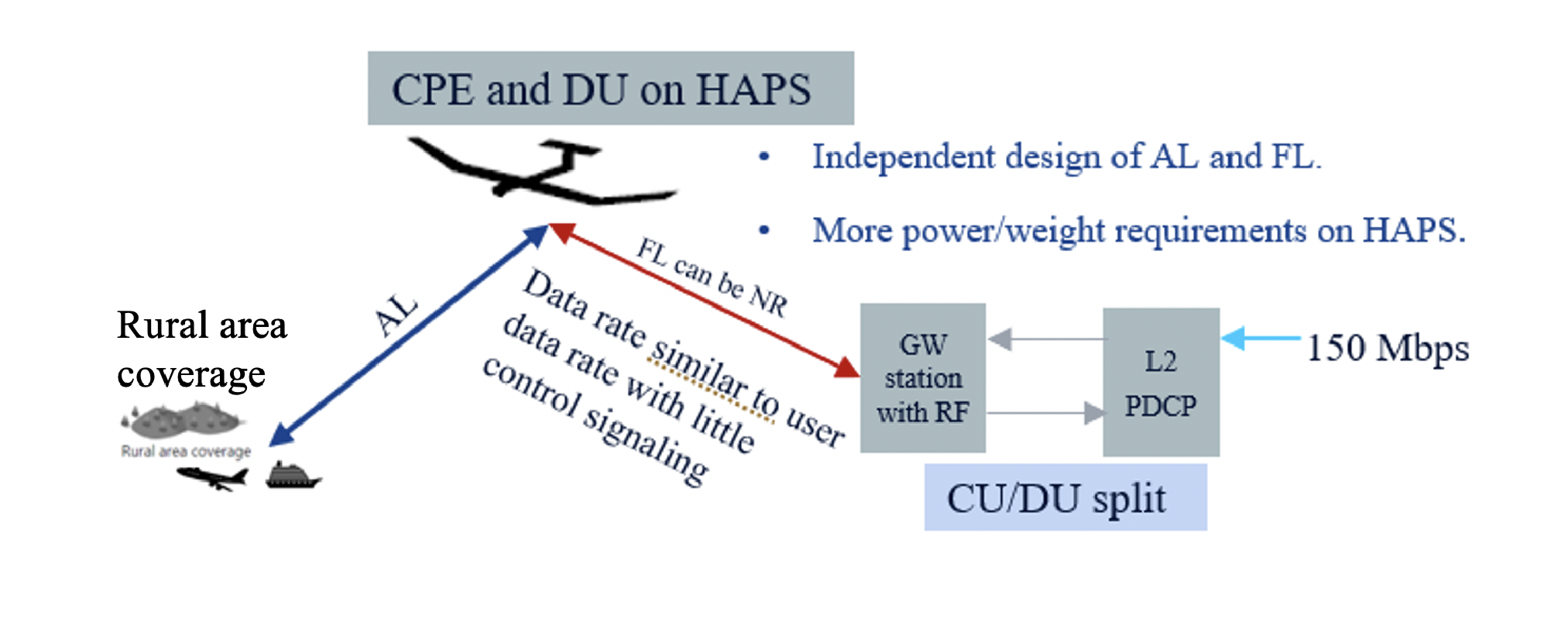}
			\caption{Re-generative HAPS architecture: HAPS as a base station.}
			\label{fig:RGarch}
				\vspace{-0.2in}
\end{figure}

The second type is regenerative (RG) architecture where the HAPS is working as a base station (e.g., radio unit (RU), distributed unit (DU) $+$ RU, or full baseband unit (BBU)) as shown in Fig. \ref{fig:RGarch}. The RG architecture HAPS can utilize \textcolor{black}{independently optimized designs} for AL and FL but it has higher power and weight requirements on HAPS. 

The specific solutions of the two architectures, including the payload and technologies requirements, are presented in Table \ref{tab:solutions}. The hardware complexity, payload requirements, and power consumption of different solutions increase from a simple repeater to a full BBU based solution. Note that the MAC layer latency (e.g., 1-2 ms) of the BP based and the RU-only based solutions (\textcolor{black}{with delay path FL$+$AL}) is higher since the base station MAC entity is located on the ground. In general, the BP architecture HAPS will be cheaper than the RG architecture HAPS system.

\section{System Model}~\label{sec:sysmodel}
A typical HAPS scenario where the HAPS is flying at an altitude of 20 km with a speed of 75-120 km/hr in a circular route of 6 km diameter to maintain a station-keeping flight pattern and provide consistent coverage on the ground \cite{frank19b}. HAPS provides data services to UEs via AL over the 4G LTE or 5G NR air interface, where lower frequency bands are preferred to provide a large coverage area. The aggregated traffic of the AL is transported by the FL, connecting the HAPS to one or more ground gateway stations (backhaul) \cite{frank19b}. In our simulations, the origin is at the cell center and a gateway station is placed 45 km away from the origin on the x-axis. 

\subsection{Single-cell and Multi-cell Cases}~\label{sec:antenna}
Both single-cell and multi-cell cases (7 cells in particular) are considered corresponding to different HAPS use cases in the simulations. A single antenna with 8 dBi antenna gain, 65\textdegree~\textcolor{black}{half power beam width (HPBW)} in both vertical and horizontal gain patterns, and a 30 dB front-to-back ratio is used to serve the single cell. A single-cell HAPS requires less hardware and power consumption, while a multi-cell HAPS utilizes a more complex antenna array to provide a larger coverage \cite{frank19b}.

\begin{figure}
	\centering
	\includegraphics[width=0.70 \linewidth]{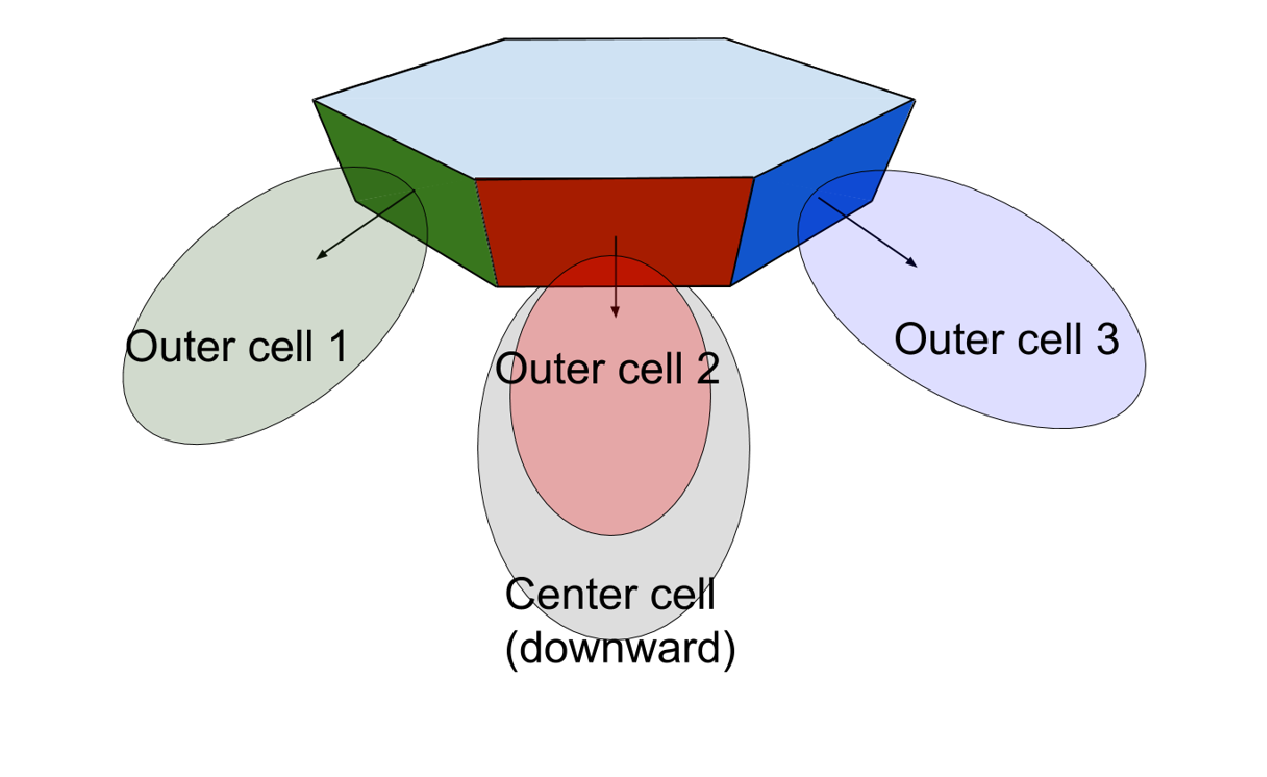}
	\caption{A hexagonal antenna array with a downward facing panel serving the center cell and six outward facing panels serving six outer cells \cite{frank19b}.}
	\label{fig:ant}
		\vspace{-0.5cm}
\end{figure}

A hexagonal antenna array structure composed of six side panels and an underneath panel facing downward to the ground as illustrated in Fig. \ref{fig:ant} is used to provide the largest possible terrestrial coverage in seven cells. The bottom panel is a $2~\text{rows} \times 2~\text{columns} \times 2~\text{polarizations}$ antenna array with boresight pointing downward, and the side panel is a $4~\text{rows} \times 2~\text{columns} \times 2~\text{polarizations}$ antenna array with $1/2\lambda$ spacing between adjacent elements, boresight pointing down 23\textdegree~from horizon. \textcolor{black}{The maximum gain of each antenna element is 5 dBi and the HPBW is 90\textdegree~in both vertical and horizontal gain patterns, with a 30 dB front-to-back ratio.} This design effectively sectorizes the large service area to seven cells, one center cell surrounded by six outer cells. 

\subsection{Access Link and Feeder Link}
LTE band 1 (DL at 2.1 GHz, UL at 1.8 GHz) with 20 MHz bandwidth and full frequency reuse in all seven cells, as well as in the single-cell case, is used for the access link (service link) in the simulation settings, and the system parameters are listed in Table \ref{tab:AL}. There are 20 uniformly distributed UEs (on the ground) for the single-cell case and 210 uniformly distributed UEs (on the ground) for the 7-cell case. All UEs are assumed to be outdoor in a rural area, and the large-scale propagation channel is modeled according to the 3GPP NTN channel model \cite{3GPP38.811}, where the LOS/NLOS status, path loss, shadow fading, and clutter loss are dependent on the elevation angle from the UE to HAPS, as has been shown in \cite{frank19a}.

The center frequency of the feeder link (the link between the HAPS and gateway station) is 3.65 GHz. The HAPS is assumed to be in unobstructed LOS link to the ground gateway station, so the propagation path loss of the feeder link is assumed to be free space path loss. The detailed system parameters of the feeder link are shown in Table \ref{tab:FL}.

\begin{table}[ht]\caption{System parameters for Access Link.}\label{tab:AL}
	\centering
	\begin{tabular}{|l|l|}
		\hline
		Environment & Outdoor Rural Marocell (RMa) area\\ \hline
	    HAPS   altitude    &  20   Km                                                                                                        \\ \hline
		 Carrier   Frequency &  DL: 2.1 GHz, UL: 1.8 GHz                                                                                      \\ \hline
		 Channel   bandwidth &  DL: 20 MHz, UL: 1 MHz                                                                                                         \\ \hline
		 Pathloss   model    &  NTN   Path Loss model \cite{3GPP38.811,frank19a}                                                                                          \\ \hline
		BS Tx power                                & 20 W per panel (43 dBm)                                                                                                               \\ \hline
		BS antenna array                           & \begin{tabular}[c]{@{}l@{}}As per Section \ref{sec:antenna} \end{tabular}                                                         \\ \hline
		BS noise figure                            & 5 dB                                                                                                                                  \\ \hline
		UE Tx power                                & 23 dBm                                                                                                                                \\ \hline
		UE Antenna Gain                            & \begin{tabular}[c]{@{}l@{}}0 dBi, omni-pattern  \end{tabular}                                                  \\ \hline
		UE noise figure                            & 7 dB                                                                                                                                  \\ \hline
		\begin{tabular}[c]{@{}l@{}}
UL resource allocation   
		\end{tabular}                  & 1 MHz per UE                                                                                                                          \\ \hline
		User distribution                          & \begin{tabular}[c]{@{}l@{}}20 UEs uniformly distributed \\for single cell cases;\\ 210 UEs uniformly distributed\\ for 7-cell cases;\end{tabular} \\ \hline
		Traffic model                              & Full buffer                                                                                                                           \\ \hline
	\end{tabular}
		\vspace{-0.5cm}
\end{table}

\begin{table}[ht]\caption{System parameters for Feeder Link.}\label{tab:FL}
	\centering
	\begin{tabular}{|l|l|}
		\hline
		Carrier frequency      & 3.65 GHz                                                               \\ \hline
		Pathloss Model & FSPL  \\ \hline
		Gateway Transmit Power & 43 dBm (20 W)                                                          \\ \hline
		Gateway Antenna        & \begin{tabular}[c]{@{}l@{}}parabolic antenna, \\ 32.3 dBi\end{tabular} \\ \hline
		Gateway Noise Figure   & 3 dB                                                                   \\ \hline
		Gateway Location       & X-axis (45 km, 0, 0)                                                   \\ \hline
		Feeder Link length     & $\sim$50 km                                                            \\ \hline
		Feeder link FSPL       & $\sim$137.7 dB                                                         \\ \hline
		HAPS repeater Gain     & 105 dB                                                                 \\ \hline
		Repeater Noise Figure  & 7 dB                                                                   \\ \hline
	\end{tabular}
\end{table}

\subsection{Repeater Model for \textcolor{black}{Bent-Pipe (BP)} HAPS}
In the design of a BP architecture system, a repeater is equipped on the HAPS to provide consistent connectivity and data service from the ground gateway station to a large area. The repeater on the BP HAPS is working as a bi-directional amplifier of RF signals in both downlink (2.1 GHz) and uplink (1.8 GHz). The repeater model specified in 3GPP TR 25.956 \cite{3GPP.25.956} is used for the BP architecture HAPS system, with 105 dB repeater gain, 7 dB repeater noise figure, and the maximum average output power of 30 dBm.

Compared to the regenerative architecture, the additional path loss of the BP architecture ($\sim$137.7 dB, the exact value depends on the instantaneous HAPS location in its flight pattern) introduced by the feeder link will be compensated by the gateway antenna gain (32.3 dBi) and the HAPS repeater gain (105 dB). Thus, using a 105 dB gain repeater in a bent-pipe HAPS system will result in approximately the same transmit power as in a regenerative HAPS system.

The noise will be amplified by the repeater by $105+7 = 112$ dB (the repeater gain plus the repeater noise figure), and then attenuated by the path loss in the access link, which will be around 121 dB to 200 dB \cite{frank19a}, resulting in a noise power level much lower than the thermal noise at the UE receivers and hence can be safely ignored.

Utilizing a higher gain repeater ($\geq 105$ dB) at HAPS will result in a higher transmit power for BP architecture than the RG architecture HAPS system, but it leads to higher power consumption and lower energy efficiency.  

\subsection{\textcolor{black}{Consumption Factor and Power-Efficiency Factor of HAPSs}}
\textcolor{black}{Consumption factor theory, as introduced in \cite{murdock2014consumption,rappaport12consumption}, is used to provide quantitative analysis and comparison of energy efficient design choices for wireless communication networks, in terms of the minimum energy consumption per bit required to achieve error-free communication \cite{murdock2014consumption}. The required energy consumption per bit can be used to evaluate the system design choices of multi-hop (repeater based bent-pipe HAPS) versus single-hop (base station based regenerative HAPS) communications.}  

\begin{figure}
	\centering
	\includegraphics[width=0.80 \linewidth]{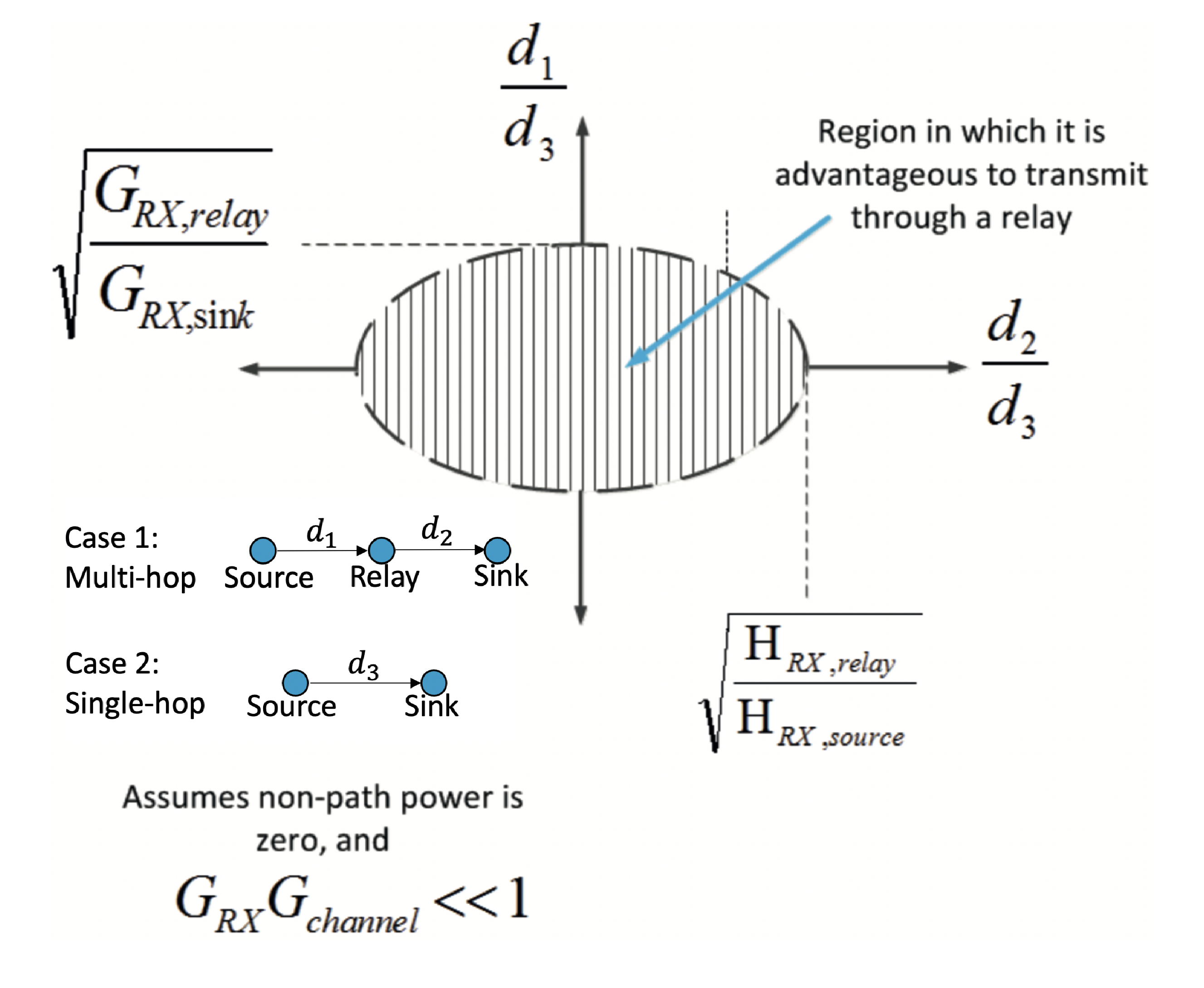}
	\caption{\textcolor{black}{It is advantageous to use a relay (repeater) provided the relay link distances are contained within the ellipse defined by \eqref{equ:ellipse} according to consumption factor theory \cite{murdock2014consumption}. This assumes free space propagation.}}
	\label{fig:Relay}
			\vspace{-0.5cm}
\end{figure}

\textcolor{black}{Fig. \ref{fig:Relay} illustrates the situation when the source and sink communicating through a relay is preferred, for the case of free-space channels, and can be expressed as (see Eq. (94) in \cite{murdock2014consumption}):
\begin{equation}
\label{equ:ellipse}
1 > \dfrac{\left( \frac{d_1}{d_3}\right)^2 }{\left( \frac{G_{RX,relay}}{G_{RX,sink}}\right) } + \dfrac{\left( \frac{d_2}{d_3}\right)^2 }{\left( \frac{H_{relay}}{H_{source}}\right)},
\end{equation}
where $d_1$, $d_2$, and $d_3$ are the distances in meters between the source to relay, relay to sink, and source to sink, respectively, as shown in Fig.\ref{fig:Relay}. $G_{RX,relay}$ and $G_{RX,sink}$ are the gain of the repeater at HAPS and gain of mobile devices in this paper (including antenna gains). $H_{relay}$ and $H_{source}$ are the power-efficiency factors of the repeater and base station at HAPS, respectively. The detailed derivations of \eqref{equ:ellipse} are provided in \cite{murdock2014consumption}. The power-efficiency factor H of a $N$-stage cascaded system can be defined as:
\begin{equation}
\label{equ:H}
H = \left\lbrace  1 + \sum_{k=1}^{N} \dfrac{1}{\prod_{i=k+1}^N G_i	\left(  \frac{1}{\eta_k} - 1 \right) }   \right\rbrace^{-1}, 
\end{equation}
where H ranges between 0 and 1, $G_i$ is the gain of the $i^{th}$ stage, $\eta_i$ is the efficiency of the $i^{th}$ signal path component. An H of 1 denotes a perfectly power efficient system that transfers all consumed energy into a rediated signal \cite{murdock2014consumption,rappaport12consumption}.}

For the bent-pipe HAPS architecture, the source is the gateway station, the relay is the repeater at HAPS, and for the regenerative HAPS architecture, the source is the base station at HAPS, with $d_1\approx 50$ km and $d_2 = d_3 > 20$ km. In this paper, since $G_{RX,relay}\gg G_{RX,sink}$ and $\left( \frac{d_1}{d_3}\right)^2 < \frac{25}{4} $, \eqref{equ:ellipse} depends on the ratio of the power efficiency factors of the relay (repeater) and source (base station). Assume a simple scenario that the repeater is a cascade of an omni-antenna $G_{R,Ant1} = 1$ to receive and transmit signals between the HAPS and gateway station, followed by a mixer with gain $G_{R,M}$ and power efficiency $\eta_{R,M}$, followed by an RF amplifier with gain $G_{R,Amp}$ and power efficiency $\eta_{R,Amp}$, and followed by an antenna to receive and transmit signals between the HAPS and UEs with gain $G_{R,Ant2}$ (assuming antennas have power efficiency of 1). The power-efficiency factor of the repeater at HAPS $H_{relay}$ is:
\begin{equation}
\label{equ:Hrelay}
\small
H_{relay} = \left\lbrace 1 + (\dfrac{1}{\eta_{R,M}} -1) + \dfrac{1}{G_{R,M}}  (\dfrac{1}{\eta_{R, Amp}} -1) \right\rbrace^{-1}. 
\end{equation}

For the regenerative architecture, assume the base station on the HAPS is a cascade of a baseband amplifier with gain $G_{B,Amp1}$ and power efficiency $\eta_{B, Amp1}$, followed by an RF mixer with gain $G_{B,M}$ and power efficiency $\eta_{B,M}$, followed by an RF amplifier with gain $G_{B,Amp2}$ and power efficiency $\eta_{B,Amp2}$, and followed by the antenna to receive and transmit between the HAPS and UEs $G_{B, Ant}$ (assuming antennas have power efficiency of 1). The power-efficiency factor of the repeater at HAPS $H_{source}$ is:
\begin{equation}
\label{equ:Hsource}
\small
\begin{split}
H_{source} = \{  1 + (\dfrac{1}{\eta_{B,Amp1}} -1) +  \dfrac{1}{G_{B,Amp1}} (\dfrac{1}{\eta_{B,M}} -1) + \\ \dfrac{1}{G_{B,Amp1} G_{B,M}} (\dfrac{1}{\eta_{B,Amp2}} -1) \} ^{-1}. 
\end{split}
\end{equation}

If the power-efficiency factor of the repeater at HAPS $H_{relay}$ is larger than the power-efficiency factor of the base station at HAPS $H_{source}$, using a repeater at HAPS (the bent-pipe architecture) is more energy-efficient than using a base station at HAPS (the regenerative architecture). 

\subsection{System Simulation Setup}
There were 12 simulation runs to compare the system performances of the two architectures of HAPSs, with 20 uniformly distributed UEs  (17 LOS, and 3 NLOS) over an area with a radius of 60 km on the ground for the single-cell case and with 210 uniformly distributed UEs (175 LOS, and 35 NLOS) over an area with a 100 km radius on the ground for the multi-cell case (7-cell). In each simulation run, the HAPS was placed in a different location on the 6km circular flight pattern centered at the origin in 30\textdegree~azimuth angle increment from the x-axis (0\textdegree~azimuth angle). The UE locations and the LOS/NLOS condition of each UE were kept the same for the 12 simulation runs with different HAPS locations. 

Note that the Doppler frequency shift induced by the HAPS velocity is minor relative to the subcarrier spacing of the OFDM waveform and can be corrected by the receiver \cite{frank19b}. Detailed simulation settings can be found in \cite{frank19b}.   

\section{Simulation results of Single cell and Multicell HAPS}~\label{sec:results}
The spectral efficiency (SE) of data transmission is an important indicator of the HAPS system performance, which usually varies across the service area depending on the UE location as well as channel conditions \cite{frank19b}. The SE of a user is the amount of data transferred to (DL) or from (UL) the user normalized by the time-frequency resources used by the system for data transfer. Suppose $M_{u}$ packets have been transmitted for user $u$, the SE $\eta_u$ is calculated as:

\begin{equation}
\eta_u = \left( \sum_{i=1}^{M_u}N_{u,i} \right)/ \left(  \sum_{i=1}^{M_u}T_{u,i}B_{u,i}\right),  
\end{equation}
where $N_{u,i}, T_{u,i}, B_{u,i}$ are respectively the number of information bits, transmission time, and allocated bandwidth for the $i$-th packet of user $u$.

\subsection{Single-cell cases}

\begin{figure}[htbp]
	\centering
		\includegraphics[width=0.8 \linewidth]{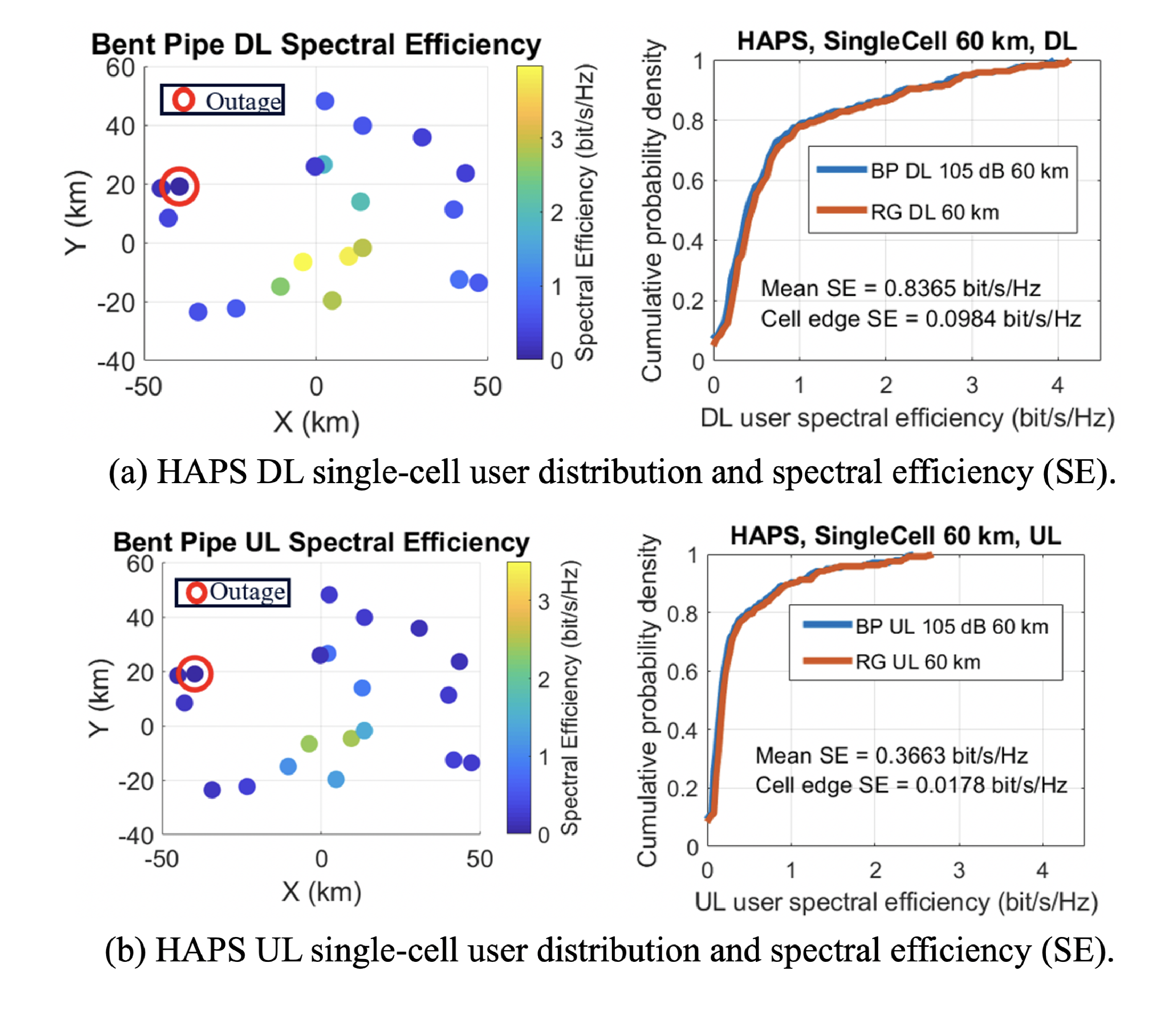}
	\caption{Single cell (60 km radius) user density distributions from system simulations with omni-directional antennas at \textcolor{black}{UEs}.}
	\label{fig:SCSE}
		\vspace{-0.1in}
\end{figure}

Fig. \ref{fig:SCSE} shows the user SE distribution in a single cell with a radius of 60 km for both downlink and uplink with all the UEs using omni-directional antennas. There is one user in outage, which is NLOS of the HAPS, out of 20 UEs, and as expected the users having a low SE tend to be located at the cell boundaries or in a NLOS shadow fade. The RG architecture and BP architecture have similar system performance as expected. A mean SE of $\sim$0.84 bit/s/Hz for DL and 0.37 bit/s/Hz for UL is observed for both architectures. The cell edge SE is defined as the average SE of the lowest 5$\%$ users, and the cell edge SE is about 0.10 bit/s/Hz for DL and 0.02 bit/s/Hz for UL. The uplink coverage is limited by the low EIRP at UEs (23 dBm). 

Table \ref{tab:SCSE} characterizes the system performance of the two HAPS architectures \textcolor{black}{in terms of} mean system SE and cell edge SE. In general, both BP HAPS with a 105 dB gain repeater and RG HAPS can provide $\sim$60 km radius coverage for one cell with DL total throughput of 17 Mbps using a 20 MHz channel bandwidth. UL mean and cell edge throughput are 366 Kbps and 18 Kbps, respectively, with a 1 MHz channel bandwidth.

\begin{table}[ht]\caption{Spectral Efficiency for a 60 km radius single cell.}\label{tab:SCSE}
	\centering
	\begin{tabular}{|l|c|c|c|c|}
		\hline
		\multirow{3}{*}{} & \multicolumn{4}{c|}{Spectral Efficiency (bit/s/Hz/cell)} \\ \cline{2-5} 
		& \multicolumn{2}{c|}{DL (20 MHz)}     & \multicolumn{2}{c|}{UL (1 MHz)}    \\ \cline{2-5} 
		& BP           & RG           & BP           & RG          \\ \hline
		Mean SE           & 0.837       & 0.841       & 0.366       & 0.370      \\ \hline
		Cell edge SE      & 0.098       & 0.105       & 0.018       & 0.019      \\ \hline
	\end{tabular}
	\vspace{-0.1in}
\end{table}

\subsection{Multi-cell cases}
Multi-cell cases with a proper design of antennas on HAPS will provide a larger coverage than the single-cell HAPS system. Multi-cell performance is studied for two implementation options: i) HAPS steers the sector beam while in its flight pattern to serve a fixed cell coverage area (beam steering, applicable to both LTE and NR) \cite{frank19b}; ii) HAPS does not steer beams, but the UE dynamically switches beam via NR beam management (beam selection) \cite{3GPP.38.214}. 

\begin{figure}[]	
	\centering
		\includegraphics[width=0.8 \linewidth]{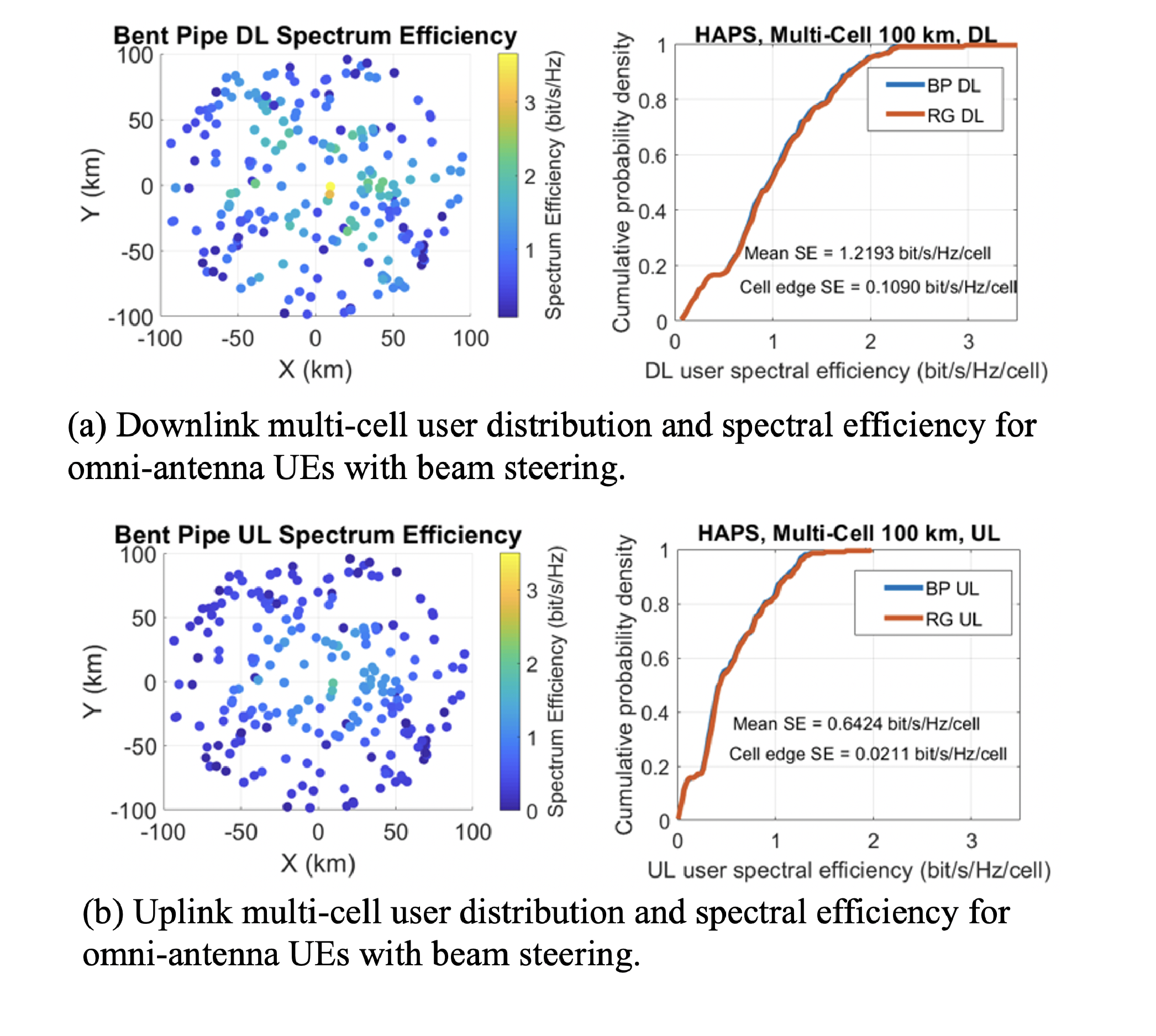}
	\caption{Multi-cell (100 km radius) user density distributions from system simulations with omni-directional antennas at UEs with beam steering on the HAPS.}
	\label{fig:MCSE}
		\vspace{-0.1in}
\end{figure}

Fig. \ref{fig:MCSE} shows the user SE distribution in a 7-cell area with a radius of 100 km for both downlink and uplink with all the UEs using omni-directional antennas, where beam steering techniques are used on the HAPS. The user SE is averaged over the 12 simulation runs with the HAPS in different locations in the flight pattern. The BP and RG architectures have similar performance in the multi-cell case, with just minor fluctuations due to feeder link path loss changes due to the HAPS movement. The downlink mean SE is 1.22 bit/s/Hz/cell and the downlink cell edge SE is about 0.15 bit/s/Hz/cell. In the uplink, the mean SE is 0.64 bit/s/Hz/cell and the cell edge SE is 0.03 bit/s/Hz/cell. Compared to the single case as shown in Fig. \ref{fig:SCSE}, using a hexagon antenna design for HAPS factorization can provide a larger coverage area and better system performance. However, the uplink coverage is still challenging in the multi-cell case. Directional high gain antennas at the UE would improve the UL SE.

\begin{figure}
	\centering
	\includegraphics[width=0.7\linewidth]{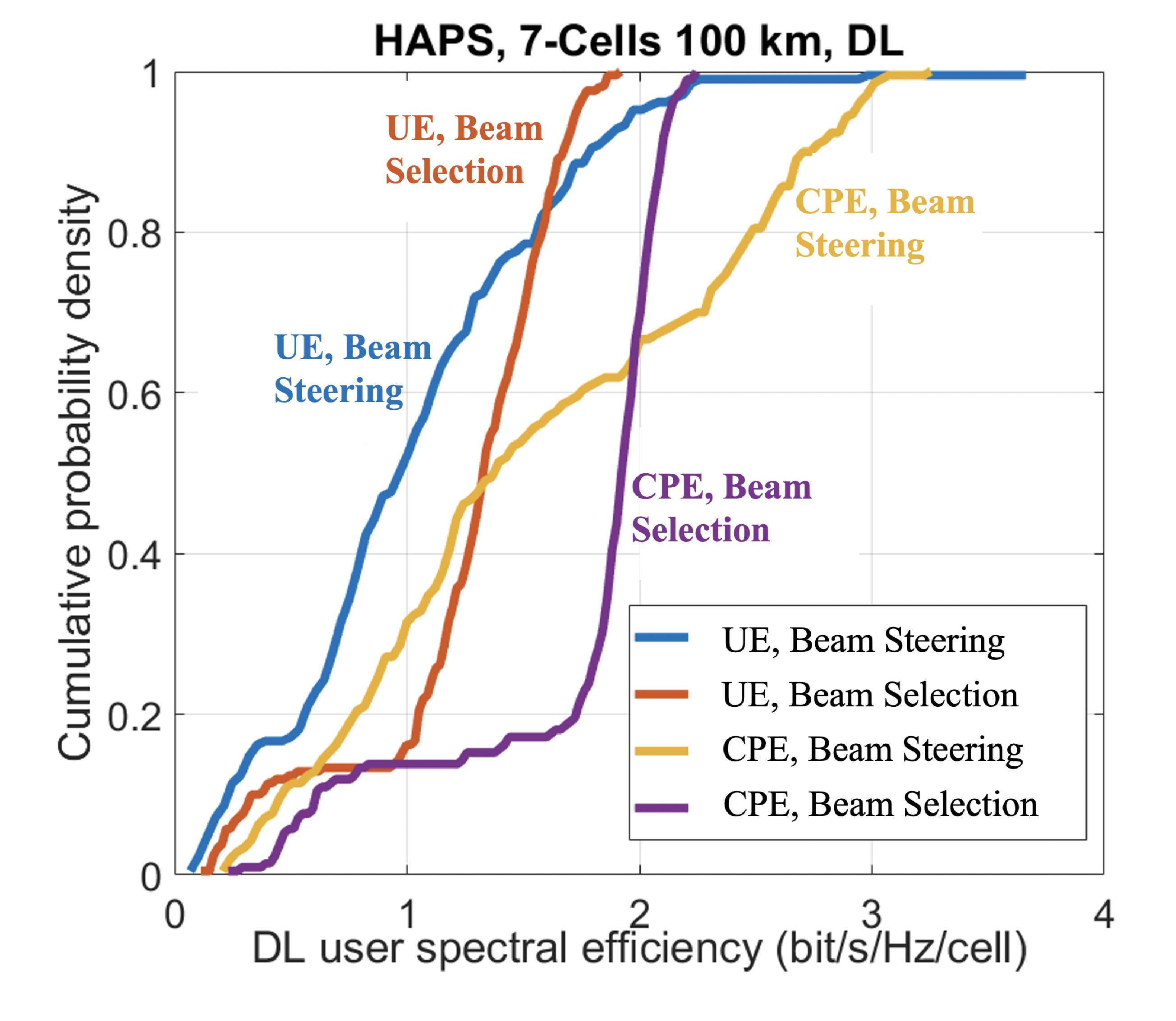}
	\caption{User \textcolor{black}{DL spectral efficiency with beam steering and beam selection for UE (omni-antenna) and CPE (directional antenna) options.}}
	\label{fig:MCSEDL}
			\vspace{-0.5cm}
\end{figure}

\begin{figure}
	\centering
	\includegraphics[width=0.7\linewidth]{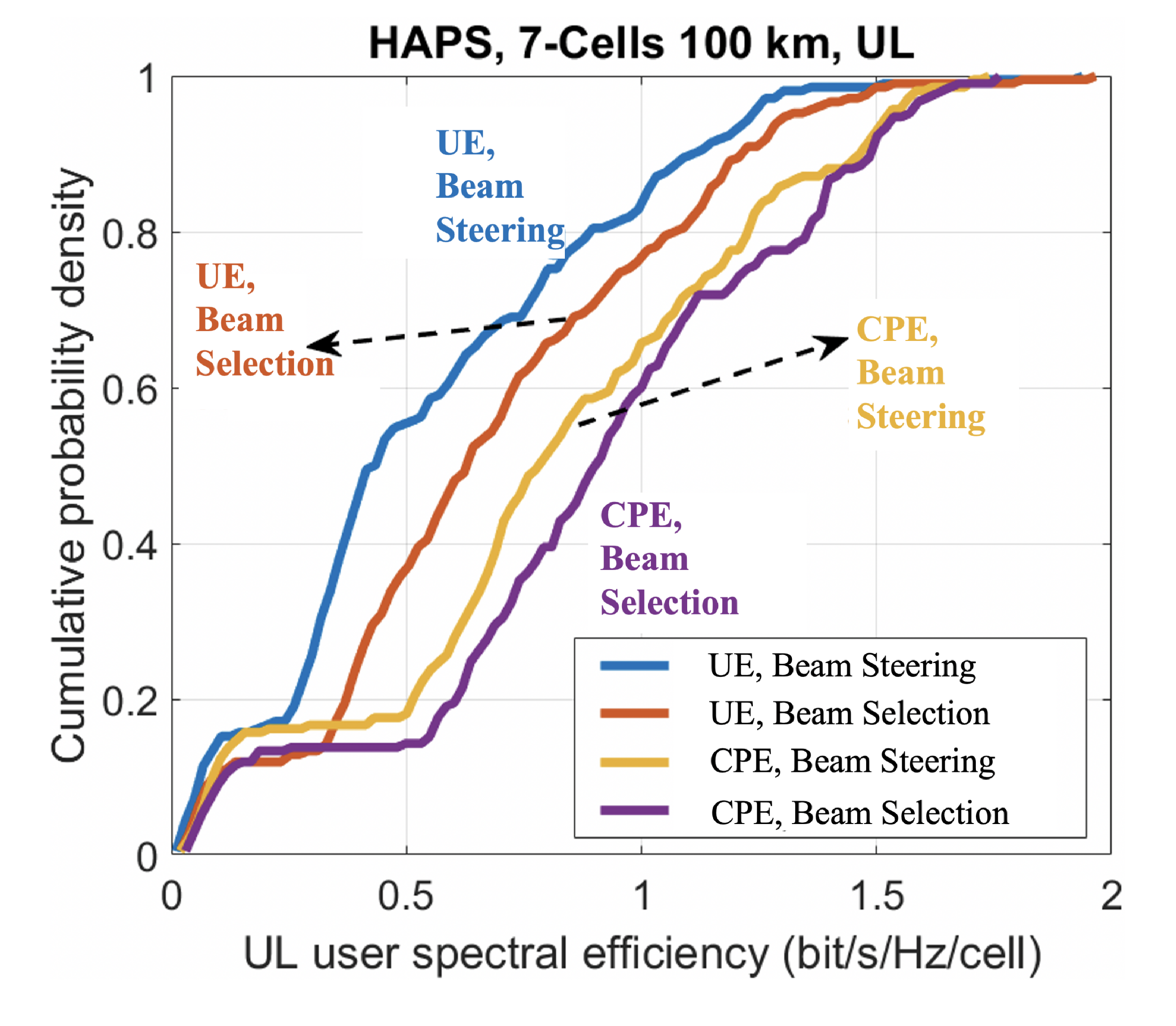}
	\caption{User \textcolor{black}{UL spectral efficiency with beam steering and beam selection for UE (omni-antenna) and CPE (directional antenna) options.}}
	\label{fig:MCSEUL}
\end{figure}

Another implementation option is to rely on the beam selection mechanism in NR based on UE's beam measurement \cite{3GPP.38.214} instead of HAPS actively steering the beam to serve a fixed cell. The system performance with the same settings and UE distributions is shown in Figs. \ref{fig:MCSEDL}, \ref{fig:MCSEUL} and Table \ref{tab:MCSE}. The downlink mean SE is 1.32 bit/s/Hz/cell and the cell edge SE is about 0.22 bit/s/Hz/cell. In the uplink, the mean SE is 0.70 bit/s/Hz/cell and the cell edge SE is 0.05 bit/s/Hz/cell. Comparison of the user SE with beam steering and beam selection are presented in Figs. \ref{fig:MCSEDL} and \ref{fig:MCSEUL}, indicating NR beam selection may provide a tangible improvement over beam steering. This is because the UE can experience a higher beamforming gain when the beam is not steered. However, the signaling overhead associated with beam measurement, reporting, and indication is not modeled in the simulations. 

We also consider the Customer Premises Equipment (CPE) with a directional antenna, assuming 12 dBi maximum antenna gain and 60\textdegree~HPBW in both vertical and horizontal gain patterns. The omni-antenna UEs are replaced with these CPE devices with the same location and fading condition in the simulations. \textcolor{black}{The directional antenna of a CPE is always} pointing to HAPS in azimuth (assume optimum orientation in azimuth), but the antenna boresight elevation angle is assumed to be 0\textdegree~(in the horizon). When beam steering \textcolor{black}{ is applied to CPE devices}, the downlink mean SE increases to 1.69 bit/s/Hz/cell and the cell edge SE increases to 0.34 bit/s/Hz/cell. The uplink mean SE increases to 0.83 bit/s/Hz/cell and the cell edge SE increases to 0.06 bit/s/Hz/cell. 

When NR beam selection \textcolor{black}{is applied to} CPE devices, the downlink mean SE increases to 1.80 bit/s/Hz/cell and the cell edge SE increases to 0.47 bit/s/Hz/cell. The uplink mean SE increases to 0.89 bit/s/Hz/cell and the cell edge SE increases to 0.06 bit/s/Hz/cell. Figs. \ref{fig:MCSEDL} and \ref{fig:MCSEUL} show that \textcolor{black}{directional antenna at the UE side} can greatly improve the throughput for both the mean and cell edge user SE, and the BP architecture and RG architecture perform similarly as expected. \textcolor{black}{For better coverage and capacity, using the NR beam selection for devices equipped with directional antennas is the preferred option.}

\begin{table}[]\caption{Spectral Efficiency for a 100 km radius multi-cell area.}~\label{tab:MCSE}
	\centering
	\begin{tabular}{|c|c|c|c|}
		\hline
		\multicolumn{4}{|c|}{7-Cell Spectral Efficiency (bit/s/Hz/cell)}                                                                                       \\ \hline
		\multicolumn{1}{|c|}{\multirow{2}{*}{Cases}}                                            & \multirow{2}{*}{} & DL (20 MHz) & UL (1 MHz) \\ \cline{3-4} 
		\multicolumn{1}{|c|}{}                                                                  &                   & BP\& RG     & BP\& RG    \\ \hline
		\multirow{2}{*}{\begin{tabular}[c]{@{}l@{}}Beam Steering \\for omni-UE \end{tabular}}        & Mean SE           & 1.219      & 0.642    \\ \cline{2-4} 
		& Cell edge SE      & 0.146      & 0.035     \\ \hline
		\multirow{2}{*}{\begin{tabular}[c]{@{}l@{}}Beam Selection \\for omni-UE\end{tabular}}     & Mean SE           & 1.323      & 0.699     \\ \cline{2-4} 
		& Cell edge SE      & 0.216      & 0.051     \\ \hline
		\multirow{2}{*}{\begin{tabular}[c]{@{}l@{}}Beam Steering \\for CPE\end{tabular}} & Mean SE           & 1.694      & 0.827   \\ \cline{2-4} 
		& Cell edge SE      & 0.339     & 0.060     \\ \hline
		\multirow{2}{*}{\begin{tabular}[c]{@{}l@{}}Beam Selection \\for CPE\end{tabular}} & Mean SE           & 1.797      & 0.889    \\ \cline{2-4} 
		& Cell edge SE      & 0.470      & 0.063     \\ \hline
	\end{tabular}
	\vspace{-0.1in}
\end{table}

\section{Conclusion}\label{sec:conclusion}
In this paper, we have presented two practical architecture designs of HAPS systems, which are capable of providing a single cell coverage of 60 km and multi-cell coverage of a 100 km radius. Both the repeater based solution and the CU/DU split based solutions are viable technical options. However, HAPS weight and power limitations and feeder link capacity dictate the practical design. \textcolor{black}{Consumption factor theory is used to provide quantitative analysis and comparison of energy efficiency of the two architectures. The situation when using the repeater based architecture is more power efficient based on consumption factor theory is presented.} While the HAPS is executing a repetitive flight pattern, fixed cell areas can be continually served by either beam steering from the HAPS or by NR beam management based on UE measurement. In the 100 km radius multi-cell case, with a 20 MHz bandwidth, DL sector throughput is 26 Mbps and DL cell edge throughput is 4 Mbps. While in the UL with a 1 MHz bandwidth, the UL mean and cell edge throughput is 0.7 Mbps and 50 Kbps, respectively. The uplink coverage is limited by UE transmit power and UE antenna gain. \textcolor{black}{Directional antennas at the UE may significantly help} improve the throughput for both DL and UL. A repeater based solution is preferred over the full BBU at HAPS \textcolor{black}{concerning} the hardware complexity and cost.  \textcolor{black}{Utilizing a high gain repeater in a bent-pipe HAPS system can result in approximately the same or even stronger transmit power as in a regenerative HAPS system, which leads to a similar or even better performance in terms of spectral efficiency even with relatively low complexity. Consumption factory theory allows quantitative comparisons of power efficiency with different architectures or designs of communication systems, providing insights for power budget analysis of future communication systems operating at data rates of hundreds of Gbps.} Terrestrial networks co-existence with HAPS systems is an essential research topic for future work. 

\bibliographystyle{IEEEtran}
\bibliography{Indoor140GHz}

\end{document}